\pdfoutput=1
\documentclass[letterpaper, 10 pt, conference]{ieeeconf}  

\IEEEoverridecommandlockouts                              

\usepackage{graphics} 
\usepackage{subcaption}
\usepackage[demo]{graphicx}
\usepackage{epsfig} 

\usepackage{amsmath} 
\usepackage{amssymb}  
\usepackage{romannum}
\usepackage{verbatim}
\usepackage[]{algorithm2e}

\usepackage{bm}
\usepackage{float}
\usepackage{multirow}
\usepackage{balance}
\usepackage[utf8]{inputenc}
\usepackage[english]{babel}

\title{\LARGE \bf
Trajectory Generation for Millimeter Scale Ferromagnetic Swimmers: Theory and Experiments
}

\author{Jaskaran Grover$^{1}$, Daniel Vedova$^{1}$, Nalini Jain$^{1}$, Patrick Vedova$^{2}$, Matthew Travers$^{1}$ and Howie Choset$^{1}$
	\thanks{$^{1}$J. Grover, D. Vedova, N. Jain, M. Travers and H. Choset are with the Robotics Institute at
		Carnegie Mellon University, Pittsburgh, PA 15213, USA.
		{\tt\small \{jaskarag@andrew.,dkv@andrew.,nalinij@andrew.,\newline mtravers@andrew.,choset@cs.\}cmu.edu}}%
	\thanks{$^{2}$P. Vedova is with the Department of Mechanical Engineering at  Rose-Hulman Institute of Technology,
		Terre Haute, IN 47803, USA.
		{\tt\small vedovaph@rose-hulman.edu}}%
}

\begin{document}
\maketitle
\thispagestyle{plain}
\pagestyle{plain}

\begin{abstract}
Microrobots have the potential to impact many areas such as microsurgery, micromanipulation and minimally invasive sensing. Due to their small size, microrobots swim in a regime that is governed by low Reynolds (\textit{Re}) number hydrodynamics.  In this paper, we consider small scale artificial swimmers that are fabricated using ferromagnetic filaments and locomote in response to time varying external magnetic fields.  We motivate the design of previously proposed control laws using tools from geometric mechanics and also demonstrate how new control laws can be synthesized to generate net translation in such swimmers. We further describe how to modify these control inputs to make the swimmers track rich trajectories in the workspace by investigating stability properties of their limit cycles in the orientation angles phase space. Following a systematic design optimization, we develop a principled approach to encode internal magnetization distributions in millimeter-scale ferromagnetic filaments. We verify and demonstrate this procedure experimentally and finally show translation, trajectory tracking and turn-in-place locomotion in these `optimal' swimmers using a Helmholtz coils setup.
\end{abstract}
\section{INTRODUCTION}

\pagenumbering{gobble}
Artificial microrobots that have characteristic lengths in the micrometer scale can have revolutionary impact for applications such as performing targeted drug delivery, microsurgery, micromanipulation and minimally invasive diagnosis. Just like microorganisms, microrobots at this scale need to swim in a low \textit{Re} number regime which warrants locomotive strategies that differ from macroscopic robots. More recently, micro-swimming \cite{koiller1996problems} has become a popular subject mainly because progress in manufacturing has made the fabrication of artificial microswimmers feasible \cite{cho2014mini,nelson2010microrobots}. 

While mechanically designing systems at these small scales is challenging, there are also unanswered questions about how to optimize, plan and control the movements of microrobots. In this paper, we explore the problem of synthesizing motion primitives for small scale magnetic swimmers that are actuated using external magnetic fields. This class of swimmers differs fundamentally from internally actuated robots where locomotion occurs in response to servo-controlled internal shape changes. However, in a low \textit{Re} regime, the minimum joint torque that is required to produce a desired angular velocity imposes a constraint on the minimum required capacity of the motors. This restriction on the minimum size of the motors and ultimately the size of the robot prevents scaling down these swimmers to micrometer dimensions. Additionally, the overall system integration with sensors, computation and actuation at the microscale becomes even more challenging \cite{peyer2013bio}. To circumvent these limitations, actuation using external time-varying magnetic fields has been proposed as an elegant and non-invasive method to drive devices without a wired connection, making it most suitable for microscopic locomotion \cite{nelson2010microrobots,cebers2016flexible}. 

In this paper, we synthesize motion primitives for articulated discrete swimmers consisting of magnetic links connected with a flexible elastomer. While mathematical models for such swimmers have been explored in \cite{alouges2015can,gutman2014simple}, we first demonstrate how tools from geometric mechanics can be used to motivate previously proposed control laws for the special case of a two-link spring-less swimmer. Secondly, we compute optimum internal magnetization distributions for such swimmers which maximize the displacement of the swimmer and also demonstrate a principled approach to encode these optimal distributions in magnetic filaments which are used to fabricate swimmers. Finally, by fabricating these optimal swimmers, we experimentally verify results from numerical simulations and demonstrate translation, turn-in-place and rectangular trajectory tracking using Helmholtz coils that generate spatially uniform time-varying magnetic fields. 
\begin{figure}
	\includegraphics[width=1\linewidth]{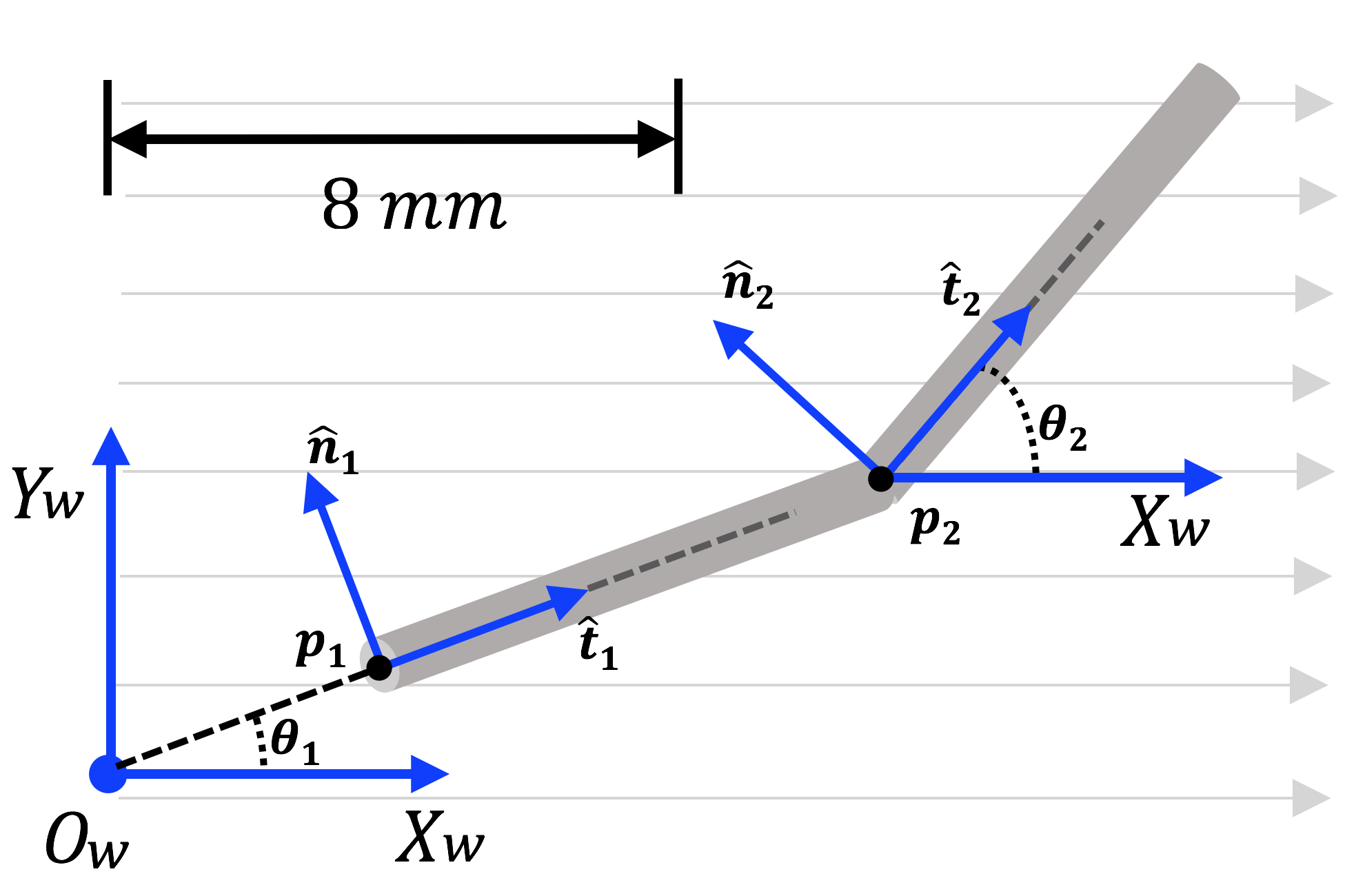}
	\caption{Top view of a two-link magnetic swimmer}
	\label{fig:twolinkcompliant}
\end{figure}
The outline of this paper is as follows. In Section \ref{priorwork} we describe prior work on existing geometric tools for low $Re$ swimmers and the state of art in magnetic swimmers. In Section \ref{Mathematical model} we describe the equations of motion for a two-link magnetic swimmer. In Section \ref{Contro laws} we describe the structure of the control laws using geometric arguments, compute optimum internal magnetizations to maximize per cycle displacement of the swimmer and also investigate stability properties of the resulting limit cycles which allows us to synthesize novel motion primitives for these swimmers. In Section  \ref{experiment trajectories}, we describe a procedure to experimentally fabricate the optimal swimmer and demonstrate translation, turn-in-place motion and rectangular trajectory tracking for the two-link swimmer in experiments. 
\section{PRIOR WORK}
\label{priorwork}
We have divided the prior work in two sections. First, we describe existing geometric and numerical tools for computing gaits for such swimmers and similar robots. We then describe existing magnetic microswimmers and their modes of propulsion.
\subsection{Geometric motion planning and optimal gait synthesis}
In the low \textit{Re} regime, viscous forces dominate swimming and inertial effects are effectively negligible. Consequently, any forward momentum gained due to internal body deformations immediately ceases to exist when the body stops changing its shape, \textit{i.e.} continuous movement requires continuous work \cite{becker2003self}. Since Purcell's initial work on the internally actuated three-link planar swimmer \cite{purcell1977life}, much of the subsequent research focuses on planning motions and computing optimal gaits for the three-link swimmer \cite{tam2007optimal}.   In addition to using numerical optimization for computing optimal gaits for these planar systems, authors in \cite{wiezel2016optimization} use the minimum principle to derive maximum-displacement gaits. Authors in \cite{hatton2011geometric} demonstrate that using visual tools derived from geometric analysis, it is possible to synthesize gaits for such a swimmer that make it move along a desired direction in the world. Authors in \cite{grover2018geometric} extended these tools to planning motions for three dimensional swimmers in such regimes. The work in \cite{gong2016simplifying} and \cite{alouges2013self} develops analytical techniques to extend gait design to articulated systems with however many links such as a snake like robot locomoting in granular media and a low-$Re$ swimmer respectively.
\subsection{Magnetic microswimmers}
In contrast to these works which mostly address motion planning for systems that are actuated internally (for e.g. via servos), there has also been significant work on inducing propulsion in such systems using actuating forces that act from outside the main body of the passive robot. Actuation using time varying magnetic fields is one such example. Such fields create a torque on uniformly magnetized links/filaments of the swimmers which respond by undergoing deformations that in turn generate propulsion. 

There are three prominent types of magnetic microswimmers that currently exist. These include a swimmer made with a rigid helical tail which propels with a corkscrew like motion in a rotating magnetic field \cite{ghosh2009controlled,peyer2013bio,sendoh2003fabrication}. The second category consists of swimmers with flexible tails connected to a passive/magnetic head as considered in \cite{khalil2014magnetosperm,honda1996micro,gao2012cargo}. Under the action of an oscillating magnetic field, the magnetic head forces the flexible tail to undulate in a non-reciprocal manner ultimately resulting in net propulsion of the swimmer. These swimmers have been analytically modeled in \cite{gadelha2013optimal,gutman2016optimizing}. The third category which is the focus of this paper, consists of articulate swimmers made with discrete links attached to each other which undergo periodic planar undulations under the influence of magnetic field with a constant and a sinusoidally oscillating component. In such swimmers, the elasticity provided by articulation inherently confers safety and locomotive efficiency to their structure at this scale. Authors in \cite{jang2015undulatory} develop such a swimmer consisting of links connected to each other with flexible hinges. By developing a lumped parameter model of such swimmers, authors in \cite{gutman2014simple} solve a parametric design optimization to determine physical parameters that maximize displacement and swimming speed.
\section{MATHEMATICAL MODEL}
\label{Mathematical model}
Fig. \ref{fig:twolinkcompliant} shows the top view of a two link swimmer which is made up of slender uniformly magnetized links and a passive joint. We assume that the swimmer is submerged in a liquid of uniform viscosity. We also assume that each link in the swimmer is ferromagnetic. This means that the internal magnetization in such a link is sufficiently strong so that a weak external magnetic field cannot distort or alter it. For each link, the longitudinal axis of the body frame is denoted by $\hat{\bm{t}}$ and the lateral axis of the body frame is denoted by $\hat{\bm{n}}$. In this representation, it is easy to see that $\hat{\bm{t}} = (\cos{\theta},\sin{\theta})$ and  $\hat{\bm{n}} = (-\sin{\theta},\cos{\theta})$ where $\theta$ is the angle between  $\hat{\bm{t}}$ and $X_{w}$ (the horizontal axis in the inertial frame). The internal magnetization of a link is denoted relative to the body frame that is rigidly attached to the link as shown in Fig. \ref{fig:twolinkcompliant}. It has components along the longitudinal axis as well as along the lateral axis of the link given by $m_th$ and $m_nh$ respectively. Here $m_t$ and $m_n$ are dimensionless numbers and $h>0$ is the internal magnetization expressed in units Am$^{-1}$. Relative to inertial frame \textit{W}, the link's magnetization can be written as
\begin{align}
\label{mag_world}
\bm{M} &=  (m_t \hat{\bm{t}}  +  m_n \hat{\bm{n}})h \nonumber \\
&= (m_th\cos{\theta}-m_nh\sin{\theta},m_th\sin{\theta}+m_nh\cos{\theta})
\end{align}
We will always assume that magnetizations are expressed relative to the world frame while still respecting that they are intrinsically defined with respect to the body frame. Given a magnetic link with volume $V$ and internal magnetization $\bm{M}$, the external magnetic field applies a torque on the link to ensure that the internal magnetic moment aligns with the direction of the external magnetic field vector. This torque is computed using 
\begin{align}
\label{mtorque}
\bm{\tau}_{m} = V \bm{M} \times \bm{B}(t) 
\end{align}
where $\bm{B}(t) = (B_x(t),B_y(t))$ denotes the $X_{w}$ and $Y_{w}$ components of the external magnetic field.

We now present the complete model of the swimmer. The configuration space of this swimmer is $Q = SE(2) \times S^1$ where the first component $SE(2)$ corresponds to the position and orientation of the body fixed frame of the first link relative to the world. The second component ($S^1$) corresponds to the orientation of the second link relative to the world. Hence the configuration variable is $\bm{q} = (x,y,\theta_1,\theta_2)=(\bm{p_1},\bm{\theta})$ where $\bm{p_1}=(x,y)$ and $\bm{\theta}=(\theta_1,\theta_2)$. The forces and torques acting on the two links are:
\begin{enumerate}
	\item $\bm{F}_{i,h}$: Hydrodynamic drag force on link $i$ expressed in the world frame ($i \in \{1,2\}$)
	\item $\bm{\tau}^{\bm{p}_m}_{i,h}$: Hydrodynamic drag torque on link $i$ relative to $\bm{p}_{m}$ expressed in the world frame ($i,m \in \{1,2\}$)
	\item $\bm{\tau}_{i,m}$: Magnetic torque on link $i$ expressed in the world frame ($i \in \{1,2\}$)
\end{enumerate}
From the assumptions of resistive force theory, it is known that the hydrodynamic forces and torques are linear in the velocity of the links. The exact expressions for these forces in terms of the velocity $\dot{\bm{q}}$ can be found in \cite{alouges2013self}. Additionally, the net force and moment on a system in a quasi-static equilibrium vanishes \textit{i.e.}
\begin{subequations}
	\label{gg}
	\begin{align}
	\bm{F}_{1,h} + \bm{F}_{2,h} &= 0  \label{eq: eom1}\\
	\bm{\tau}^{\bm{p}_1}_{1,h} + \bm{\tau}^{\bm{p}_1}_{2,h} + \bm{\tau}_{1,m} + \bm{\tau}_{2,m}&= 0 \label{eq: eom2}\\
\bm{\tau}^{\bm{p}_2}_{2,h} + \bm{\tau}_{2,m}&= 0 \label{eq: eom3}
\end{align}
\end{subequations}
After substituting the expressions for the forces and torques and rearranging, we can rewrite Eqs. \ref{eq: eom1}-\ref{eq: eom3} in the form of a control affine system. The control input to the system is defined by the spatial magnetic fields \textit{i.e.} $\bm{u} = (B_{x}(t),B_{y}(t))$ and the state of the system is $\bm{q}(t)$. 
\begin{align}
\label{mag_Swimmer_dynamics}
\dot{\bm{q}} &= \bm{g}_1(\bm{q})B_x(t)  + \bm{g}_2(\bm{q})B_y(t) \nonumber  \\
&= \bm{G}(\bm{q})\bm{u} \nonumber \\
&\bm{q} \in \mathbb{R}^4 \\ \nonumber
&B_x,B_y:[0,T] \longrightarrow \mathbb{R} \nonumber
\end{align}
We define a vector $\bm{m}=(m^1_t,m^2_t,m^1_n,m^2_n)$ that denotes the internal magnetizations of the links (the subscript indicates tangential/normal while the superscript indicates the link index). To highlight the dependence of the matrix $\bm{G}(\bm{q})$ on $\bm{m}$, we will explicitly denote it as $\bm{G}(\bm{q},\bm{m})$.
\section{CONTROL LAWS }
\label{Contro laws}
\begin{figure}
	\centering
	\includegraphics[width=1.0\linewidth]{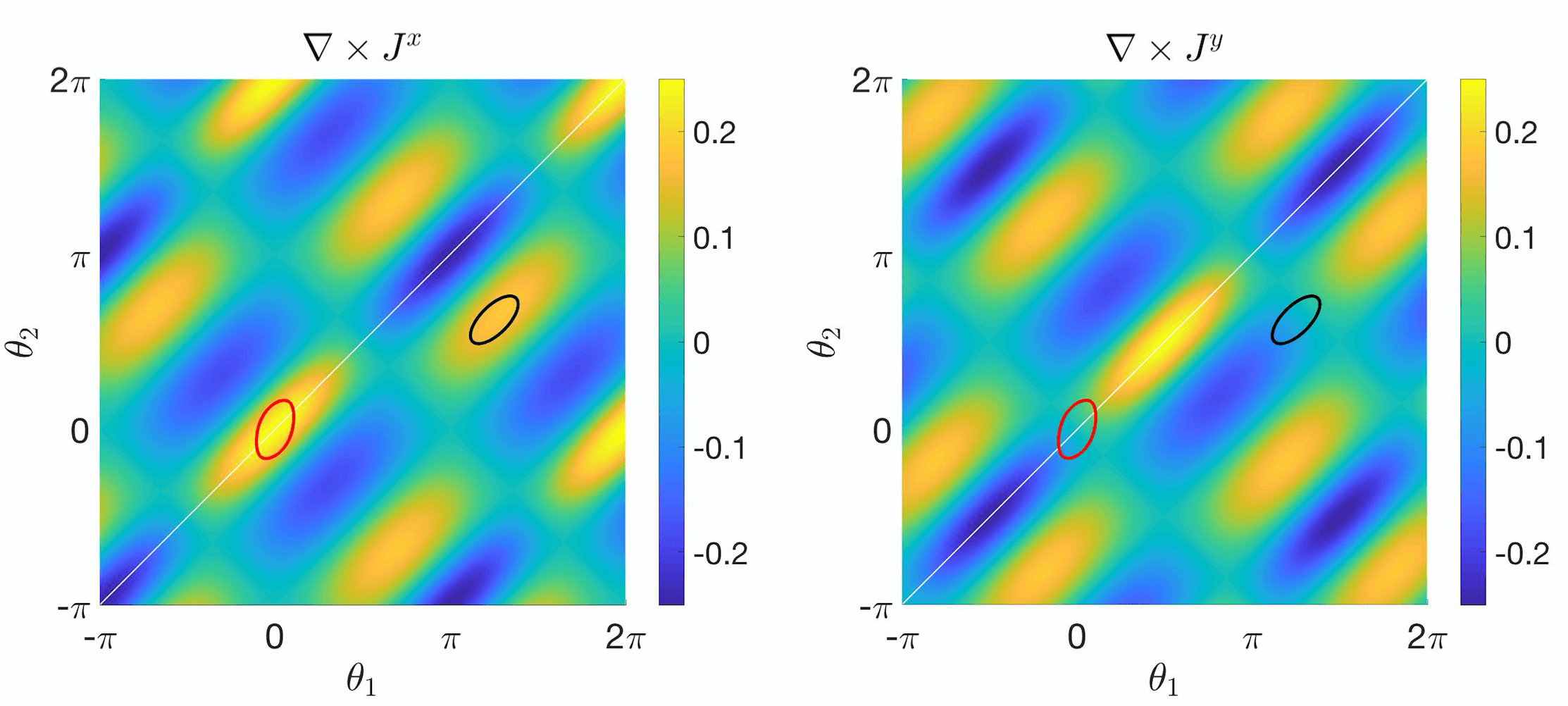}
	\caption{\mbox{curl} $\bm{J}$ computed over $[-\pi,2\pi] \times [-\pi,2\pi]$}
	\label{fig:geometricheightfunction}
\end{figure}

\begin{figure}
	\centering
	\includegraphics[width=1\linewidth]{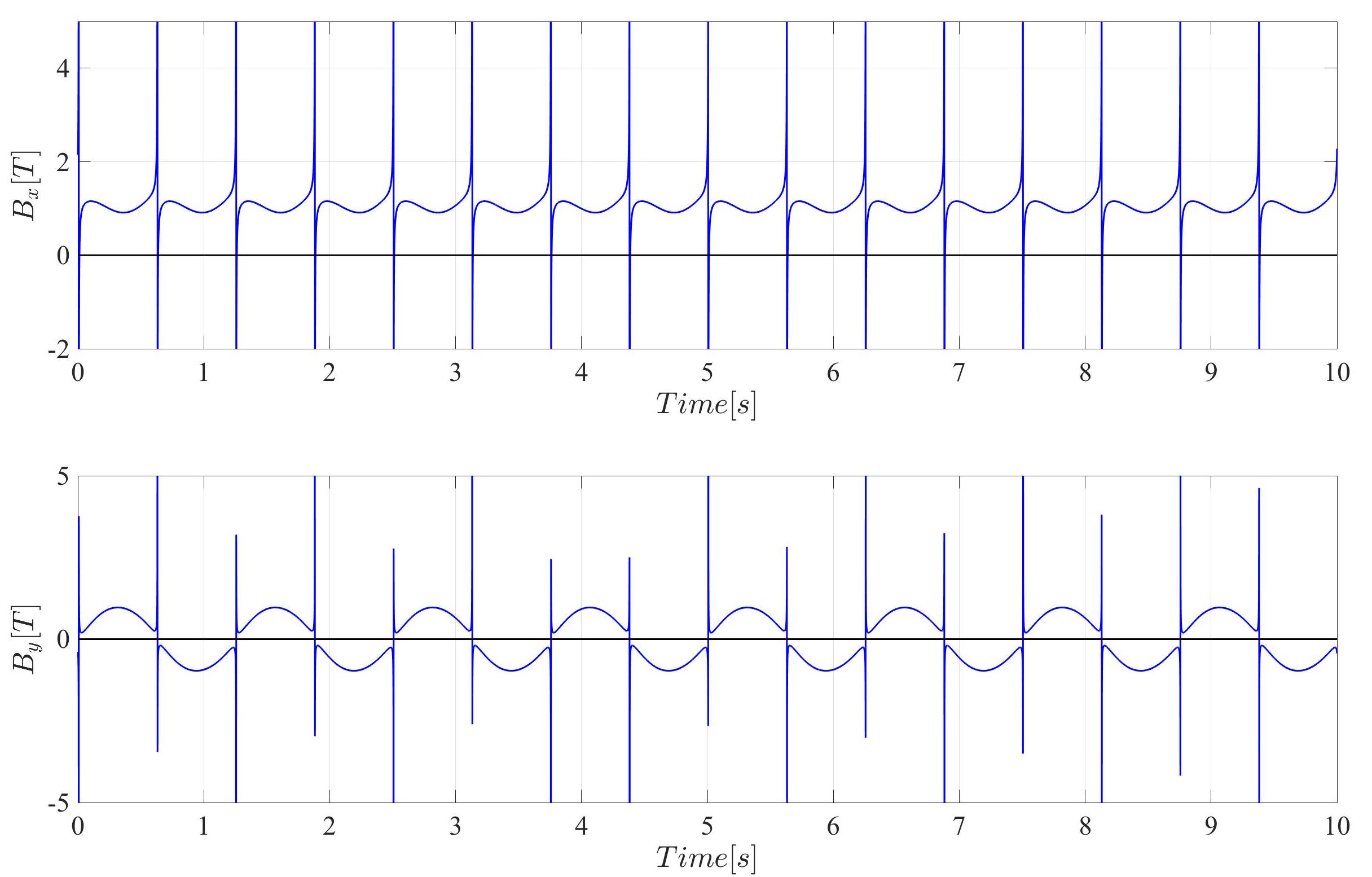}
	\caption{Control input for the red loop computed from Eq. \ref{eq: desired_control} }
	\label{fig:controlfrominversion}
\end{figure}

In the previous section, we formulated the equations of motion for a two-link magnetic swimmer. Based on those equations, we can explore the behavior of the swimmer's motion as a function of different types of control inputs. We will only consider the case where the swimmer does not have a spring, as driftlessness of the dynamics is essential to the ensuing development. Consider the differential equations governing the dynamics of the two link swimmer:
\begin{align}
\label{driftless_mag_Swimmer_dynamics_repeat}
\dot{\bm{q}} &= \bm{g}_1(\bm{q},\bm{m})B_x(t)  + \bm{g}_2(\bm{q},\bm{m})B_y(t) \nonumber  \\
&= \bm{G}(\bm{q},\bm{m})\bm{u}
\end{align}
where $\bm{q} = (\bm{p},\bm{\theta})$, \mbox{$\bm{G}(\bm{q},\bm{m}) = \left[\bm{g}_1(\bm{q},\bm{m}),\bm{g}_2(\bm{q},\bm{m})\right]$} and \mbox{$\bm{u}(t)=(B_x(t),B_y(t))$}. 
Since the magnetic field is spatially uniform, the instantaneous $(x,y)$ position coordinates of the swimmer do not effect its motion. The only state variables that influence the dynamics are the orientation of the swimmer's links in the inertial frame. Therefore, the dynamics of the position variables \textit{i.e.} $\dot{\bm{p}} = (\dot{x},\dot{y})$ depend exclusively on the orientation variables $ \bm{\theta} = (\theta_1,\theta_2)$. Similarly, the dynamics of the orientation variables $\dot{\bm{\theta}}$  depend exclusively on $\bm{\theta}$. Hence, we can break Eq. \ref{driftless_mag_Swimmer_dynamics_repeat} into two separate sub-systems as follows:
\begin{align}
\dot{\bm{q}} &= \bm{G}(\bm{q},\bm{m})\bm{u} =\bm{G}(\bm{\theta},\bm{m})\bm{u} \nonumber \\
\implies \dot{\bm{p}} &= \bm{P}(\bm{\theta},\bm{m})\bm{u}  \label{eq:geo_eq1}\\
\dot{\bm{\theta}} &= \bm{H}(\bm{\theta},\bm{m})\bm{u} \label{eq:geo_eq2}
\end{align}
where  $\bm{P}(\bm{\theta},\bm{m}) \in \mathbb{R}^{2 \times 2}$ and $\bm{H}(\bm{\theta},\bm{m}) \in \mathbb{R}^{2 \times 2}$ . Assuming $\bm{H}(\bm{\theta},\bm{m})$ is invertible on $[-2\pi,2\pi] \times [-2\pi,2\pi]$, we can compute $\bm{u}$ from Eq. \ref{eq:geo_eq2} and substitute in Eq. \ref{eq:geo_eq1} as follows:
\begin{subequations}
	\label{geo_mag_development_equations}
	\begin{align}
	\bm{u} &= \bm{H}^{-1}(\bm{\theta},\bm{m}) \dot{\bm{\theta}} \label{eq: control_for_mag_geometry}\\
	\implies \dot{\bm{p}} &= \bm{P}(\bm{\theta},\bm{m}) \bm{H}^{-1}(\bm{\theta},\bm{m}) \dot{\bm{\theta}}  \\
	\implies  \dot{\bm{p}}  &= \bm{J}(\bm{\theta},\bm{m}) \dot{\bm{\theta}} \label{eq : mag_kre}
	\end{align}
\end{subequations}
where $\bm{J}(\bm{\theta},\bm{m}) = \bm{P}(\bm{\theta},\bm{m}) \bm{H}^{-1}(\bm{\theta},\bm{m})$. Note that Eq. \ref{eq : mag_kre} is in a form similar to the \textit{Kinematic Reconstruction Equation}
\begin{align}
	\label{recon}
	\bm{\xi} = -A(\bm{\alpha})\dot{\bm{\alpha}}
\end{align}
where $ A(\bm{\alpha})\in \mathbb{R}^{ 3 \times 2}$ is known as the local form of a connection. It maps shape velocities to body velocities: $A(\bm{\alpha}): T_{\alpha_1} S^1 \times T_{\alpha_2}S^1 \longrightarrow \mathfrak{se}(2)$.  In the literature on geometric theory of swimming \cite{kelly1995geometric}, this equation has been used to synthesize motion primitives for swimmers that are internally actuated \textit{i.e.} where it is possible to command any values of $\bm{\alpha}(t)$. On the contrary, note that in Eq. \ref{eq : mag_kre} the variables $(\theta_1,\theta_2)$ refer to the orientation of the swimmer relative to the world. Additionally, the left hand side  of Eq. \ref{eq : mag_kre} also involves velocities referenced relative to the inertial frame as opposed to the body velocities $\bm{\xi}$ expressed in the body frame. Hence, we cannot model this system with a principal fiber bundle structure. Nevertheless, assuming for the moment that we can fully and independently control $(\theta_1,\theta_2)$, it is possible to compute the total displacement over a cyclic change in $(\theta_1,\theta_2)$ as follows
\begin{align}
\bm{p}(T) &= 
\int_{0}^{T}\bm{p}(t)dt \nonumber \\
&= \int_{0}^{T}\bm{J}(\bm{\theta}(t),\bm{m})\dot{\bm{\theta}}(t) dt \nonumber \\
&= \int_{\gamma}\bm{J}(\bm{\theta},\bm{m}) d\bm{\theta} \nonumber  \\
&= \iint_\mathcal{S}\nabla \times \bm{J} \mbox{  } d\theta_1 d\theta_2. 
\label{eq: mag_stokes}
\end{align}
In Eq. \ref{eq: mag_stokes}, we have used Stokes' theorem\footnote{Differently from the work in \cite{alouges2015can}, we use Stokes' theorem for magnetic swimmers without springs and  visualize the effect of the limit cycles over the full orientation space using curvature function plots in Fig. \ref{fig:geometricheightfunction}.  Secondly, we do not require a small-angle approximation and instead illustrate the effect of the full swing limit cycle on displacement directly.} to simplify the problem of computing line integral of the rows of $\bm{J}(\bm{\theta},\bm{m})$ along $\gamma$, to computing volume integrals defined over $\mathcal{S}$. 
We can now compute loops in the $(\theta_1,\theta_2)$ space to get a desired displacement of the swimmer in the world. We plot the $x$ and $y$ components of $ \mbox{curl }\bm{J}$ in Fig. \ref{fig:geometricheightfunction}. To synthesize a motion primitive for translation in the $X_{w}$ direction of the world frame, we query regions in Fig. \ref{fig:geometricheightfunction} in the orientation angle phase space which enclose a net non-zero volume in $x$ component and zero volume in the $y$ component. One such candidate loop is highlighted in red in Fig. \ref{fig:geometricheightfunction} and is parametrized as 
\begin{subequations}
	\label{eq: desired_thetas_limit}
	\begin{align}
	\theta_1^d(t) &= 0.35\sin{(\omega t-1.817)}   \label{eq: desired_theta_1_limit}\\ 
	\theta_2^d(t) &= 0.53\sin{(\omega t -0.7186)}   \label{eq: desired_theta_2_limit}
	\end{align}
\end{subequations}
Using this parametrization, it is possible to compute a control law using Eq. \ref{eq: control_for_mag_geometry} which is computed point-wise in time:
\begin{align}
\label{eq: desired_control}
\bm{u}(t) = \bm{H}^{-1}(\theta_1^d(t),\theta_2^d(t))\left[\begin{matrix}
\dot{\theta^d_1} \\
\dot{\theta^d_2}
\end{matrix}\right]
\end{align}
The resulting control law is plotted in Fig. \ref{fig:controlfrominversion}.
\subsection{Relation to previously proposed control laws}
\label{relation to previous control laws}
\begin{figure}
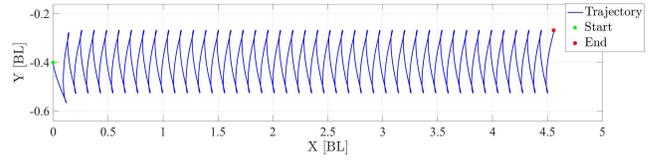

	\centering
	\includegraphics[width=01\linewidth]{{{Images/Mag_Swimmer_Pics/simpletranslate}}}
	\caption{Translation trajectory using $B_x = 1, B_y = \sin{\omega t}$}
	\label{fig:xi0}
\end{figure}

 Note that the control input in Fig. \ref{fig:controlfrominversion} exhibits a discontinuity which happens instantaneously whenever $\theta_1(t)=\theta_2(t)$. This singularity is the result of the swimmer admitting an instantaneous straightened configuration and is by no means a locked singularity. We low-pass filter the control inputs to remove the singularity and interpolate and normalize them by their amplitude. This gives
\begin{align}
\label{final_control_input}
B_x(t) &= 1 \nonumber \\
B_y(t) &= \sin{\omega t}
\end{align}
This control input matches exactly with the ones proposed in \cite{gutman2014simple,alouges2015can}. Using this control, we numerically simulate the dynamics of the system \textit{i.e.} Eq. \ref{mag_Swimmer_dynamics} and plot the resulting trajectory of the swimmer as shown in Fig. \ref{fig:xi0}.

\subsection{Novel motion primitives}
\label{novel primitives}
\begin{figure}
	\centering
	\includegraphics[width=01\linewidth]{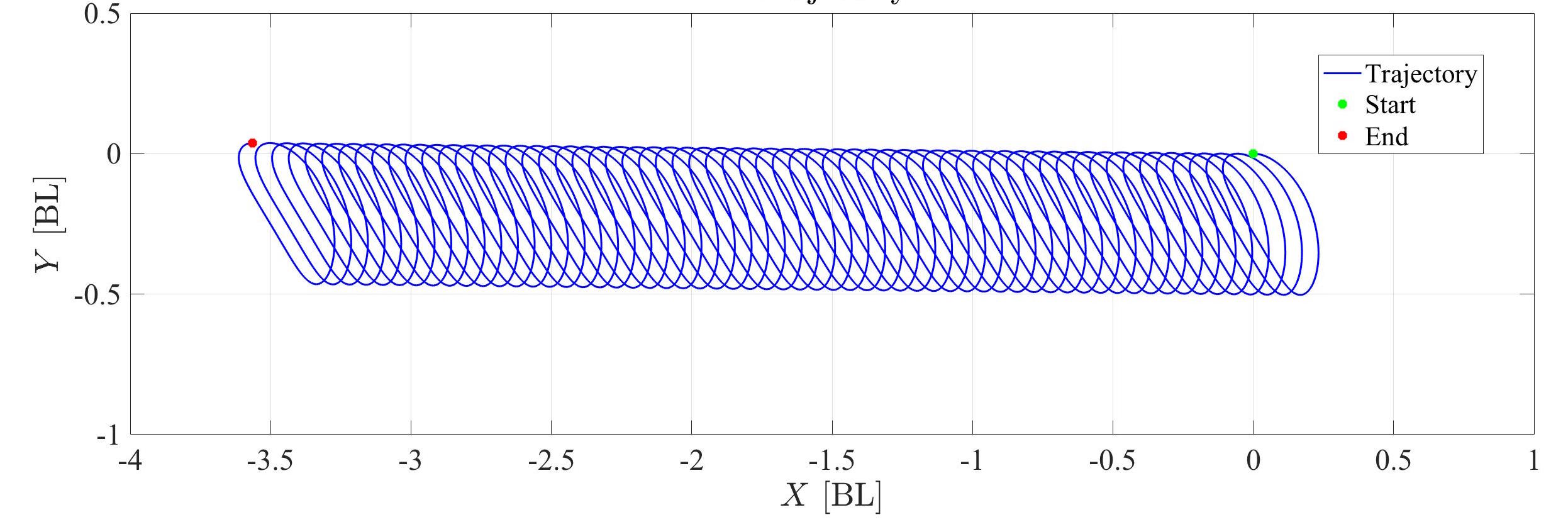}
	\caption{Translation trajectory corresponding to the magnetic fields computed using Eq. \ref{eq: desired_control} for \ref{eq: desired_theta_1}-\ref{eq: desired_theta_2}}
	\label{fig:geometrictranslation}
\end{figure}

We extend this tool to synthesize new control inputs that also result in translation along $X_{w}$ axis in the world. To that end, consider a time parametrized loop in the $(\theta_1,\theta_2)$ defined below
\begin{subequations}
	\label{eq: desired_thetas}
	\begin{align}
	\theta_1^d(t) = 0.5\cos{\omega t} \cos{\frac{\pi}{4}} - 0.25\sin{\omega t}\sin{\frac{\pi}{4}} + \frac{5\pi}{4} \label{eq: desired_theta_1}\\ 
	\theta_2^d(t) = 0.5\cos{\omega t} \sin{\frac{\pi}{4}} + 0.25\sin{\omega t}\cos{\frac{\pi}{4}} + \frac{3\pi}{4} \label{eq: desired_theta_2}
	\end{align}
\end{subequations}
This loop is depicted in black in Fig. \ref{fig:geometricheightfunction}. Using this parametrization, we compute $\bm{u}(t)$ using Eq. \ref{eq: desired_control} and simulate the system dynamics using this control. The resulting displacement of the swimmer is shown in Fig. \ref{fig:geometrictranslation} and it indeed undergoes translation along ${X_{w}}$ axis.

\subsection{Optimal Swimmer}
Given the control input \ref{final_control_input}, we pose the following optimization problem to compute internal magnetization distributions ($m^1_t=c_1$ and $m^2_t=c_2$) that maximize the per cycle displacement of the swimmer (from $t=0$ to $t=T=\frac{2 \pi}{\omega}$)
\begin{equation*}
\begin{aligned}
& \underset{c_1,c_2}{\text{maximize}}
& & \vert x(T) \vert \\
& \text{subject to}
& & \dot{\bm{q}} = \bm{G}(\bm{q},\bm{m})\bm{u}  \\
& &\bm{u}(t) &= (1,\sin{\omega t}) \\
& &\bm{q}(0)&=\bm{0}\\
\end{aligned}
\end{equation*}
We use brute force numerical simulations to solve this optimization problem. Based on the numerical results of this optimization, we selected $\frac{c_2}{c_1}=2$ which represents a swimmer where link 2 has twice as strong an internal magnetic moment as link 1. \footnote{We restrict to $\frac{c_2}{c_1}=2$ because experimental fabrication of arbitrarily large $\frac{c_2}{c_1}$ is not feasible}. 
\subsection{Limit cycle stability}
\label{Limit cycles}
\begin{figure}
	\centering
	\includegraphics[width=1\linewidth]{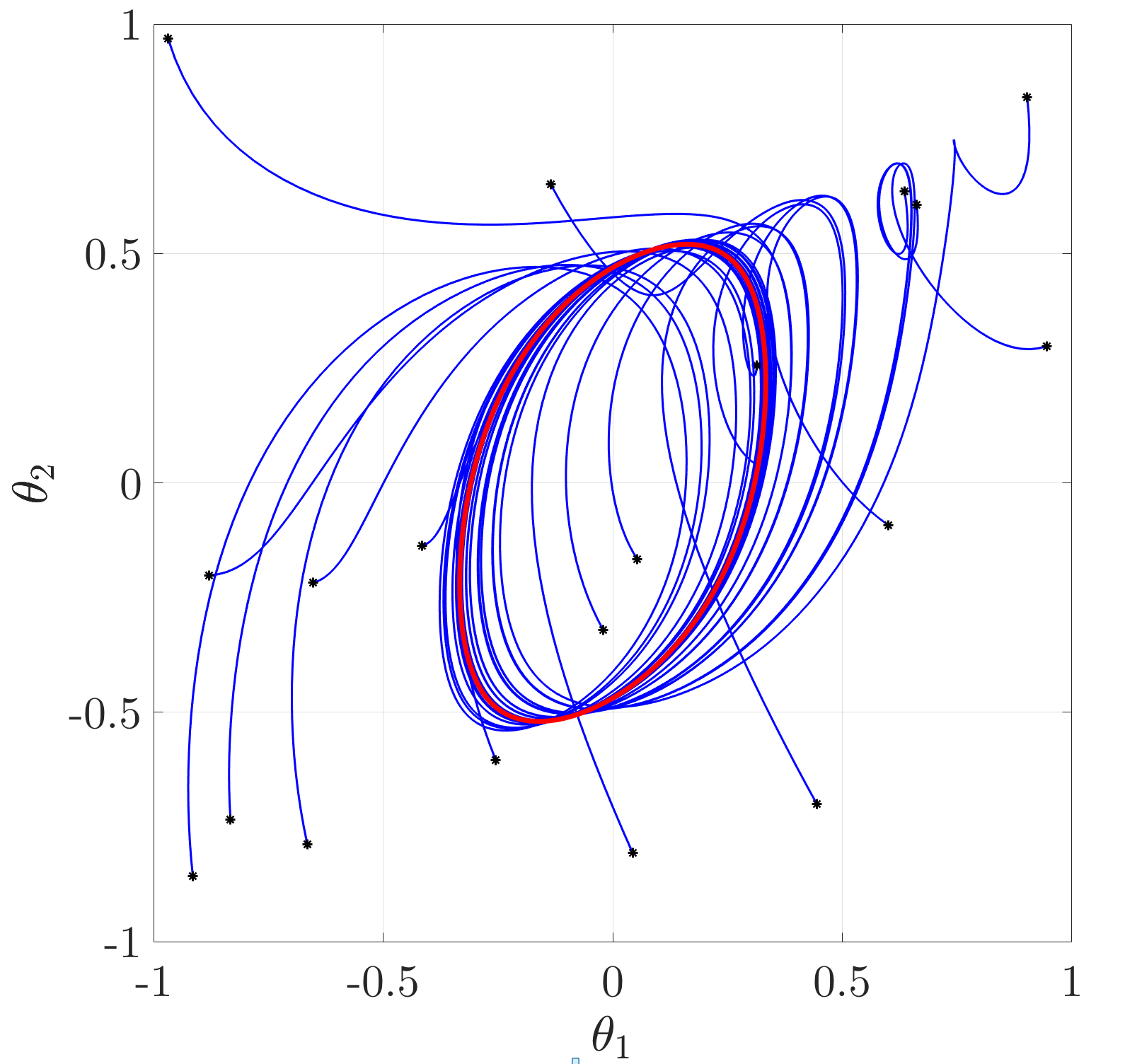}
	\caption{Limit cycles in $(\theta_1,\theta_2)$ from different initial orientations using \mbox{$B_x = 1, B_y = \sin{\omega t}$ }}
	\label{fig:limitcycles}
\end{figure}

\vspace{-0.6cm}
From numerical simulations, we observe that the steady state motion of the swimmer using the control input $\bm{u}(t)=(1,\sin{\omega t})$ is independent of the swimmer's initial orientation in the world. To illustrate this point, we plot the flows of the orientation components of the system state starting from different initial conditions in Fig. \ref{fig:limitcycles}.  This figure demonstrates that all these flows converge to the same limit cycle in $(\theta_1,\theta_2)$ space as highlighted in red. We now prove that this limit cycle is locally asymptotically stable. This proof is numerical and demonstrated by computing fixed points of the Stroboscopic map and eigenvalues of the jacobian of this map w.r.t the the orientation variables $\frac{\partial S}{\partial \bm{\theta}}$. Due to space limitations, we omit this proof here and refer the reader to \cite{granados2016notes} for a systematic algorithm to numerically compute these eigenvalues. Following this approach for our system, we find that the resulting eigenvalues of $\frac{\partial S}{\partial \bm{\theta}}$ are $\{0.0603,0.000083147\} <1$. Hence the limit cycle highlighted in red in Fig. \ref{fig:limitcycles} is locally asymptotically stable.
Note that the large basin of attraction in the limit cycle demonstrates that it is possible to align the swimmer with the external magnetic field even when the initial orientation of individual links of the swimmer is off by as much as $\pi$ radians relative to the direction of the constant component of the external field. Hence, we can exploit this fact to make the swimmer translate along a given trajectory in the world by suitably switching the direction of the constant and the oscillating component. We will use this property to make the swimmer track a rectangular trajectory, the arc of a circle and make it turn-in-place.
\section{EXPERIMENTAL VALIDATION}
\begin{figure}
	\centering
	\includegraphics[width=1\linewidth]{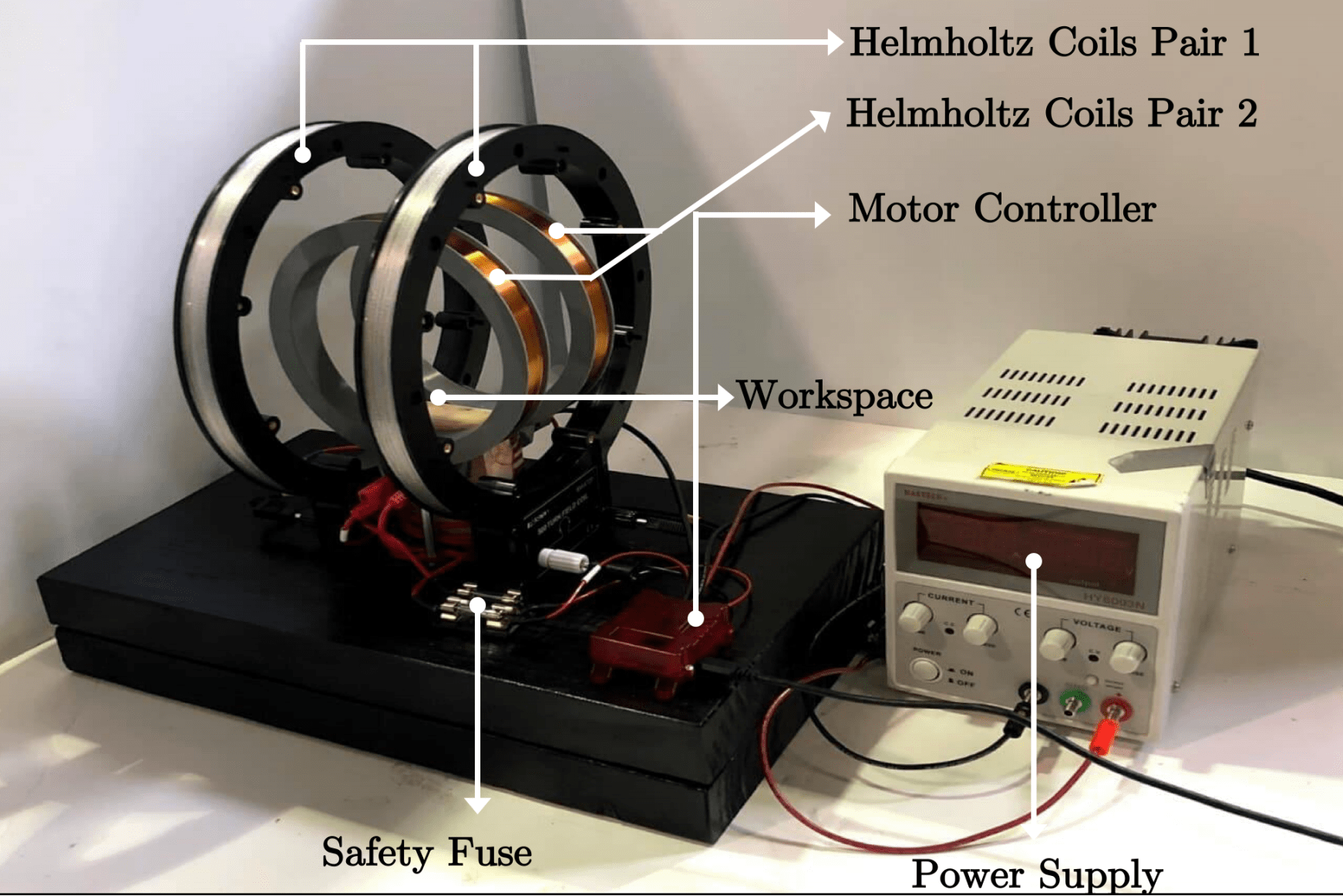}
	\caption{Experimental setup used to generate planar magnetic fields}
	\label{fig:expsetup}
\end{figure}

\begin{figure*}%
	\centering
	\begin{subfigure}{0.6\columnwidth}
	\includegraphics[width=01\linewidth]{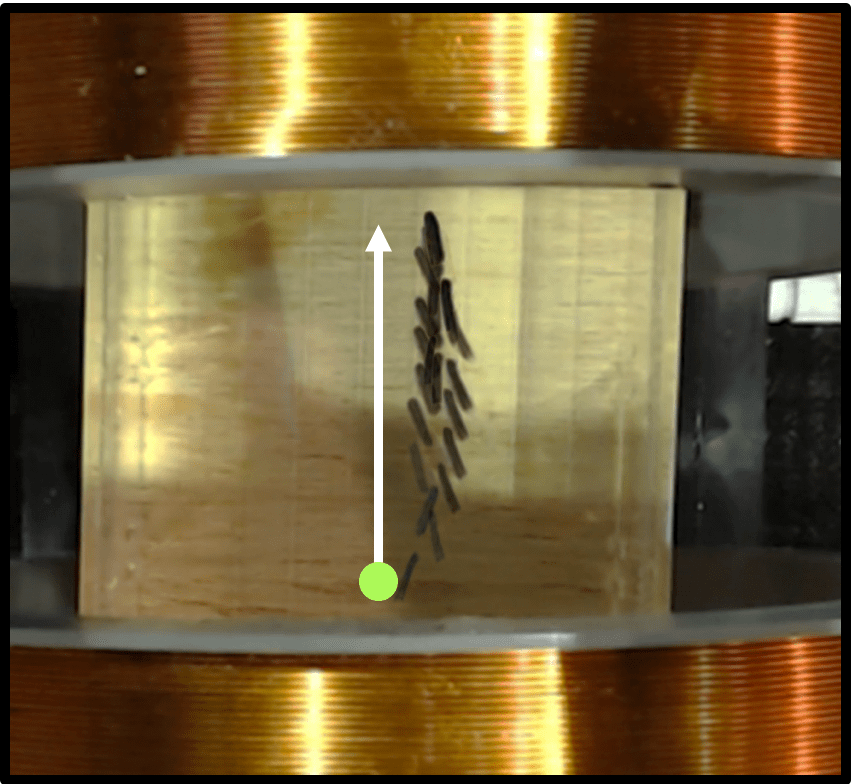}
		\caption{Translation in the two-link swimmer}%
		\label{subfiga}%
	\end{subfigure}\hfill%
	\begin{subfigure}{0.6\columnwidth}
\includegraphics[width=1\linewidth]{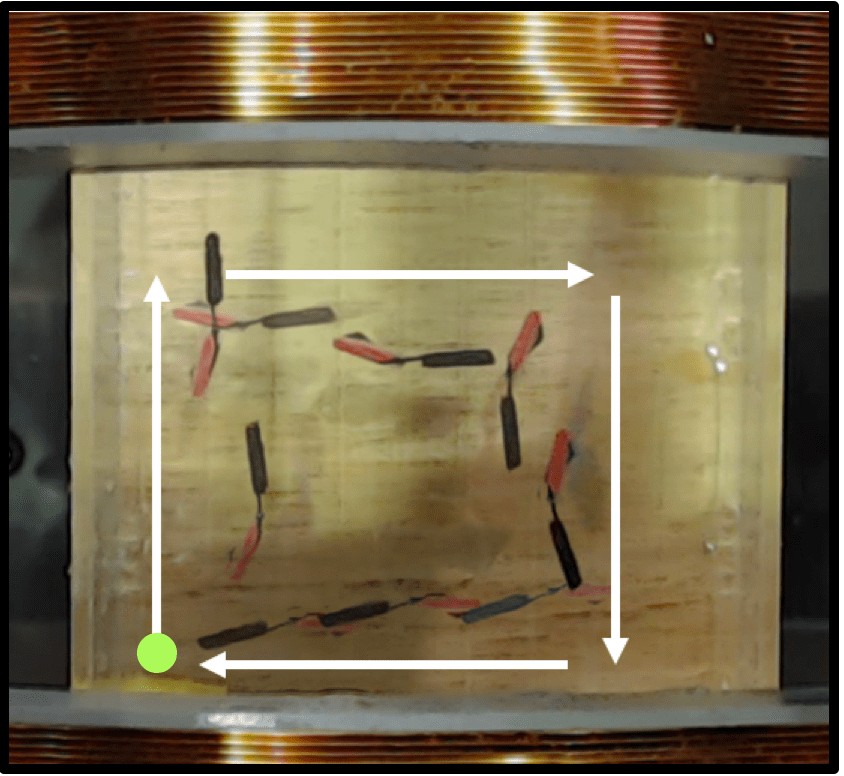}
		\caption{Rectangular Trajectory Following}%
		\label{subfigb}%
	\end{subfigure}\hfill%
	\begin{subfigure}{0.6\columnwidth}
	\includegraphics[width=1\linewidth]{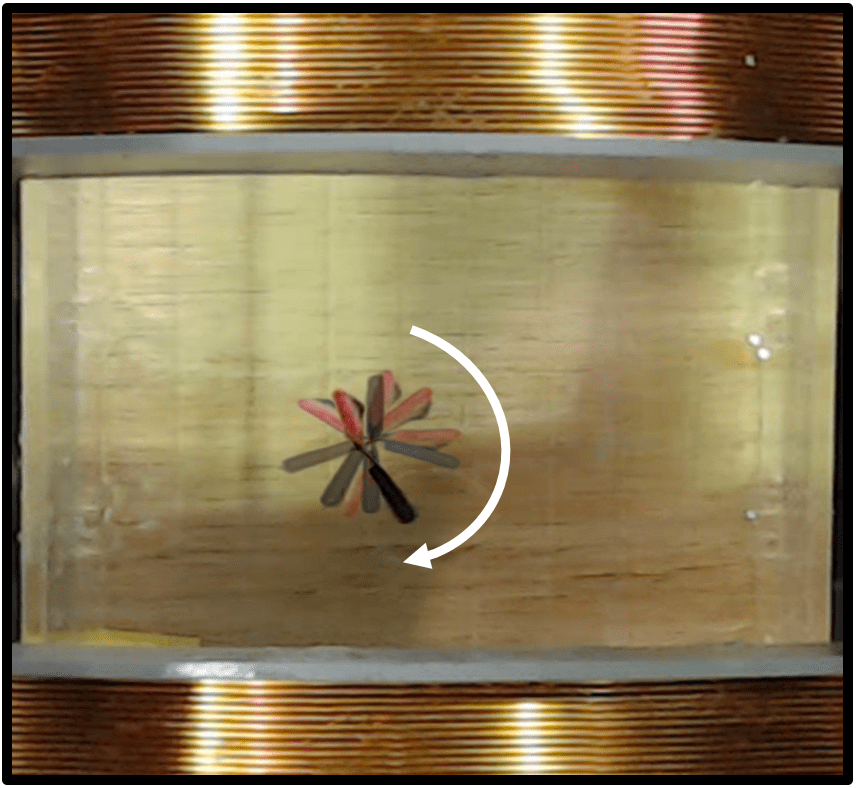}
		\caption{Turning in place}%
		\label{subfigc}%
	\end{subfigure}%
	\caption{Translation, rectangular trajectory tracking and turning in place.}
	\label{figabc}
\end{figure*}

\label{experiment trajectories}

\subsection{Setup}
Fig. \ref{fig:expsetup} shows the experimental setup to validate the trajectories of the swimmer derived from simulations. We use two pairs of Helmholtz coils and a \textit{Roboclaw} motor driver to generate a spatially uniform magnetic field ($\sim$ 200 Gauss) in $X_{w}$ and $Y_{w}$ direction in the workspace. Feedback information about the motion of the swimmer is provided by a \textit{Logitech C920} camera. The length of the swimmer is 1cm and it is submerged in glycerin to simulate low-$Re$ motion.
\subsection{Swimmer fabrication}

\label{swimmer_fabrication}
\begin{figure}
	\centering
	\includegraphics[width=1\linewidth]{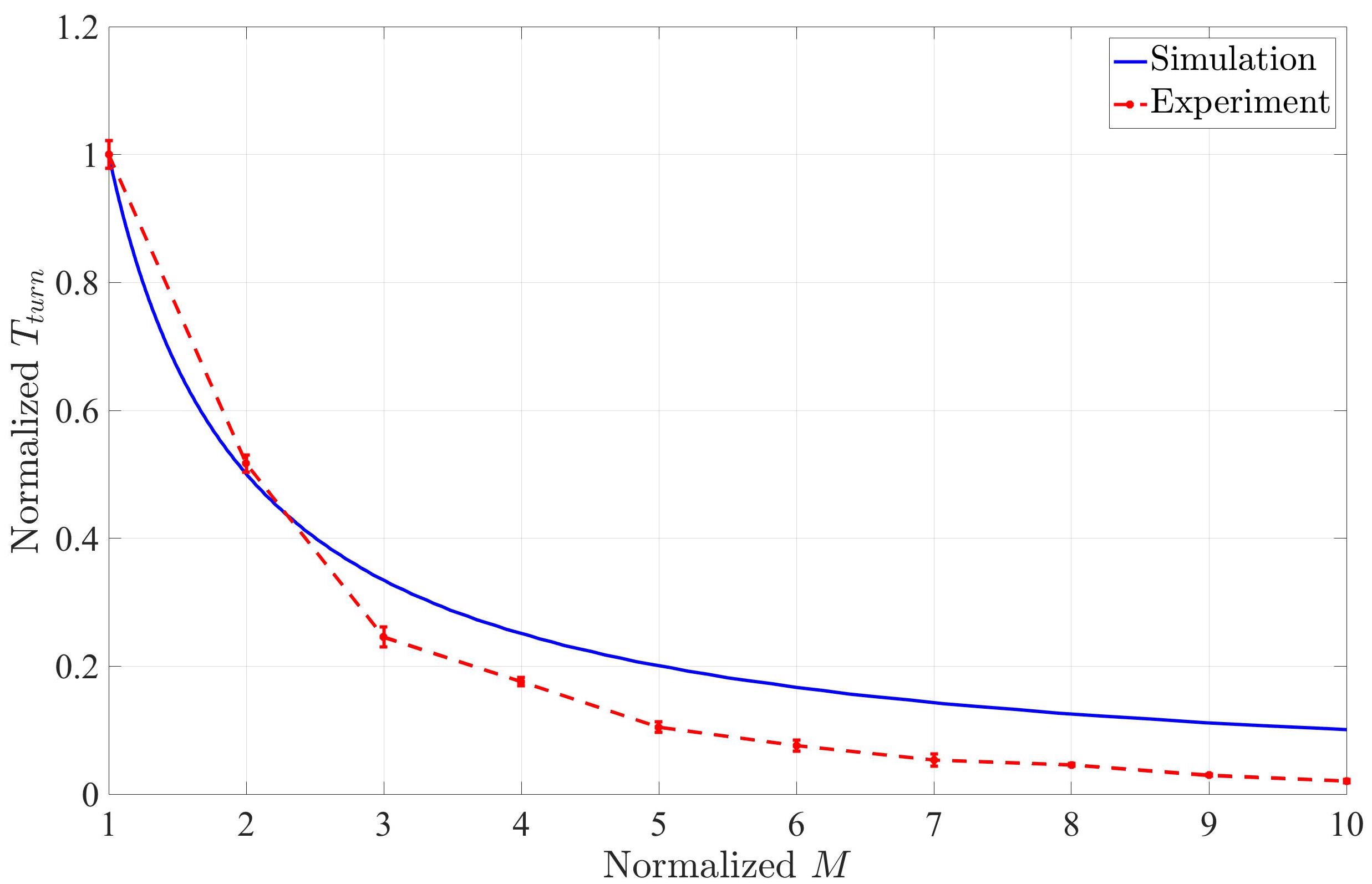}
	\caption{Turning time as a function of internal magnetization}
	\label{fig:turningtime}
\end{figure}

The full swimmer is fabricated by connecting two ferromagnetic links with non-magnetic elastomer.  To encode the desired magnetization profile in the swimmer, we use a pair of strong neodymium magnets that generate strong magnetizing fields of different values depending on inter-magnet distance. Each link is suspended in the region between these magnets and left for 24 hours to allow sufficient time for magnetic domains to align unidirectionally. Once aligned, we join the links with an elastomer. We assume that the strength of the magnetization induced inside the link $M_{induced}$ is directly proportional to the strength of the field $B_{magnetizing}$ between the two magnets. For simulations, we decided to encode a ratio where link 2 was twice as strong a magnet as link 1. To realize this ratio experimentally, we conduct a turning time test. In this test, we investigate the time a single link submerged in glycerin takes to turn and align itself with the magnetic field as a function of the internal magnetization when the initial orientation of the link is perpendicular to the applied magnetic field. From simulations, we find that the time taken to turn is inversely related to the internal magnetization,. To verify this experimentally, we magnetize several links, measure the local magnetizing field on the link and measure the time to turn by $\frac{\pi}{2}$ for these links using libraries in $OpenCV$. As Fig. \ref{fig:turningtime} shows, the experimental data matches well with simulations, and indeed this test allows us to compare magnetizations between different links and can be used to fabricate a swimmer with 1:2 magnetization.

\subsection{Translation}
We now present our results from applying $\bm{u}_{trans}(t) = B_0(1,\sin{\omega t})$ to the swimmer. For these experiments, we used $B_0 = 30 \mbox{ Gauss}$ and $\omega = 2\pi \mbox{ rad/s}$. Fig. \ref{subfiga} depicts snapshots of the translation of the tail of the two-link swimmer as a function of time. 
\subsection{Rectangular trajectory following}
To track a rectangular trajectory \footnote{\texttt{https://youtu.be/-iMNkjYARu8}}, we synthesize the switching time instants $(t_1,t_2,t_3,t_4)$ and control inputs $\bm{u}(t)$ at which the constant and the oscillating components of the magnetic field turn counter clockwise by $\frac{\pi}{2}$ radians (Eq. \ref{switch}, here $R_{\alpha}$ is a 2D rotation matrix). Whenever this happens, the swimmer exhibits a transient response which eventually diminishes and steady translation of the swimmer is obtained along the constant component. Fig. \ref{subfigb} depicts snapshots of the two-link swimmer following rectangular trajectory.
\begin{equation}
\label{switch}
\bm{u}(t) =  \begin{cases} 
\bm{u}_{trans}(t) & 0\leq t \leq t_1 \\
R_{\frac{\pi}{2}}\bm{u}_{trans}(t)& t_1< t \leq t_2 \\
R_{\pi}\bm{u}_{trans}(t) & t_2< t \leq t_3 \\
R_{\frac{3\pi}{2}}\bm{u}_{trans}(t)& t_3 < t \leq t_4 \\
\end{cases}
\end{equation}

%



\subsection{Turn-in-place}
Following a similar approach as before, we now modulate the magnetic field in a continuous fashion by composing it with a frequency component slower than the original actuating frequency $\omega$. This causes the swimmer to continuously rotate as the instantaneous magnetic field vector on average also rotates\footnote{\texttt{https://youtu.be/b9wSUphyhm8}}. Fig. \ref{subfigc} demonstrates an example of turning in place motion for the two link swimmer. For this experiment, we used $\bm{u}_{rot}= R_{\omega_{slow}t}\bm{u}_{trans}(t)$. The oscillating field frequency was $\omega=2\pi \mbox{ rad/s}$ and $\omega_{slow} = \frac{\omega}{10}$.  

\section{CONCLUSIONS}
In this paper, we explored motion planning methods for a two-link magnetic swimmer which is actuated using external magnetic fields. We also conducted experiments to verify translation, trajectory tracking and turn in place motions in these swimmers. From our experiments, we noticed that the swimmer did not perfectly follow the rectangular trajectory which could be attributed to the swimmer's closeness to walls that reflects momentum and also to the inherent passive dynamics of the swimmer's compliant design which prevent the swimmer's links from beating about the symmetry axis.  In future, we would like to design swimmers which have minimal passive dynamics, make efficient swimmers which devote greater energy to translate forward as well use closed loop control to correct deviations in the swimmer's motion. We would also like to extend the geometric control synthesis tools to designing control inputs for $N$-link magnetic swimmers. Although prior work has demonstrated that control inputs similar to the ones considered here generate net translation in $N$ link swimmers, we would like to derive a differential-geometric motivation for using these inputs. Finally, we would like to investigate stability properties of the limit cycles as a function of link magnetizations and frequency. Computing this relation can be used to inform the procedure to program these profiles and make swimmers that can selectively respond to specific frequencies of actuating fields, opening up the possibility of making articulated mechanisms for cargo-towing for example.

\addtolength{\textheight}{-10cm}   
\balance

\addcontentsline{toc}{section}{References}
\bibliography{References}
\bibliographystyle{ieeetr}

\end{document}